\documentclass[twocolumn]{jpsj3}
\usepackage{txfonts}
\title{A Phenomenological Theory of Loop-Current Phases}

\author{Shimpei Goto and Susumu Kurihara}
\inst{Department of Physics, Waseda University, Shinjuku, Tokyo 169-8555, Japan} %\\

\abst{A phenomenological theory of the loop-current and loop-spin-current phases is proposed. In order to investigate the stability of these phases, 
a Ginzburg-Landau-Wilson type action is constructed as a functional of the orbital magnetization. 
From the analysis of this action based on the Landau theory and
momentum-shell one-loop renormalization group theory, 
it is found that the loop-current and loop-spin-current phases are stable if a certain interaction between the orbital magnetizations is sufficiently large. 
Moreover, these phases are likely to be stable in systems with large orbital susceptibility, for example, Dirac electron systems.}

\begin{document}
\maketitle

%introduction
Orbital motions of itinerant electrons in condensed matter usually give rise to weak diamagnetism.
As the magnetization from the localized spins, the magnetization from the orbital motions breaks the time-reversal symmetry 
and can be treated as an order parameter, which lowers the symmetry of the system. 
The {\it loop-current} phase is a quantum phase characterized by the orbital magnetization caused by a local electric current forming a loop.

This phase has been introduced in the studies of cuprates in order to understand the physical mechanism of pseudo gap phenomena \cite{PhysRevB.37.3774, PhysRevB.55.14554, PhysRevLett.83.3538, PhysRevB.73.155113}.
Like other phases emerging in cuprates, the loop-current phase is expected to be induced by Coulomb interactions.
For instance, a theoretical study based on a three-band Hubbard model has shown that the loop-current phase is stable in some parameter region \cite{PhysRevB.55.14554}.
Furthermore, in the half-filled Hubbard model on a square lattice, a numerical study by the variational cluster approximation (VCA) \cite{PhysRevB.85.125117} has indicated that this phase has metastable characters; the free energy of this phase is lower than that of the trivial phase, but higher than that of another ordered phase.

Interestingly, the topologically nontrivial states, such as the Chern insulators \cite{PhysRevLett.49.405}, may have some relations to the loop-current phase.
As Hofstadter has shown\cite{PhysRevB.14.2239}, the single-particle spectrum of the non-interacting electron on a square lattice with the magnetic field perpendicular to the system has characteristic pattern which is often referred to as the Hofstadter butterfly. 
Calculating the Chern number $C$ for each band in the butterfly, one obtains a nonvanishing value of $C$, which indicates the nontrivial topological character of the system.
In these calculations, the effect of the magnetic filed is included as the Peierls phase, which forces itinerant electrons to make a circular motion on the average.
Thus, the electronic states of the loop-current phase is expected to be similar to those in the butterfly.

Indeed, the numerical study based on the VCA on a honeycomb lattice\cite{ourpapar} has shown that the loop-current phase induced by the Coulomb interaction has a nontrivial Chern number as expected. 
In this study, authors have also included another current phase called the {\it loop-spin-current} phase, 
which is characterized by the circulating flow of spins unaccompanied by charge currents. 
In the loop-spin-current phase on a honeycomb lattice, the system has a nontrivial $Z_2$ invariants \cite{PhysRevLett.95.146802, PhysRevB.74.195312}, which assure different topological characters from those assured by a nontrivial Chern number. 
According to the topological field theory\cite{PhysRevB.78.195424, PhysRevLett.105.256803}, 
the nontrivial $Z_2$ invariants mean the existence of the $\mib{E} \cdot \mib{B}$ term, which indicates that the system exhibits the magnetoelectric response.

As described above, the loop-current and loop-spin-current phases are related to recent hot topics in condensed matter physics.
Nonetheless, general features of these phases are not known well.
In particular, unified descriptions for the loop-current phases are required in order to find a criterion for the emergence of these phases.
For these phases induced by an inter-site interaction, a unified picture is given by the mean-field analysis of the free energy\cite{PhysRevB.88.115143}.
However, this picture cannot be applied directly to the phases induced by a local interaction, such as the on-site Hubbard interaction,
since loop-current orders do not emerge from a mean-field decoupling of the on-site interaction.

Consequently, we construct a simple phenomenological description for loop-current and loop-spin-current phases induced by a local interaction.
Since orbital motions of itinerant electrons induce the orbital magnetization, we treat this magnetization as the order parameters. 
Furthermore, the spin degrees of freedom is introduced to the orbital magnetization in order to describe the loop-spin-current phase.
A local interaction is included as couplings between the local orbital magnetization from up spin and that from down spin.

In this letter, we analyze  the stability of the local orbital magnetization phenomenologically.
From the analysis based on the Landau theory, we find that both the loop-current and the loop-spin-current phases emerge from the interaction between the orbital magnetizations.
The momentum-shell renormalization group (RG) theory also shows that those phases are stable against sufficiently small spatial fluctuations.
Moreover, it is shown that these phases are expected to emerge in systems with large orbital susceptibility. 

%method
Our analyses are based on the following
time-reversal-symmetric Ginzburg-Landau-Wilson type action treating the orbital
magnetization as the order parameters:
\begin{eqnarray}
\label{GLW}
S_{\mathrm{GLW}}[\phi_\sigma]\!\!\!\!\!&=& \!\!\!\!\!\!
\int \mathrm{d}^d
\mib{x}
\sum_{\sigma}\!\left \{ \frac{1}{2} \left[ \nabla \phi_{\sigma}(\mib{x}) \right ]^2 +
\frac{m}{2}\phi^2_{\sigma}(\mib{x}) + \frac{g}{4!}\phi^4_{\sigma}(\mib{x}) \right \}  \nonumber \\
 &&\!\!\!\!\!\!+ \int \mathrm{d}^d
\mib{x}J\phi_{\uparrow}(\mib{x})\phi_{\downarrow}(\mib{x}).
\end{eqnarray}
Here, $\phi_{\sigma}(\mib{x})$ denotes the orbital magnetization from the
electron at point $\mib{x}$ with spin $\sigma$, $d$ is spatial dimension, and $m,J,$ and $u$ are the
parameters; $m$ represents the energy cost for generating magnetization,
$g$ is the quartic local interaction, and 
$J$ is the bilinear local interaction between the magnetizations. 
For the thermodynamic stability, $g$ should be positive. 
We assume that $m$ is positive.
In other words, we assume that generating the orbital magnetization requires the energy cost corresponding to the kinetic energy of the electron.
As is well known, the positive $m$ indicates the absence of the ordered phase in the isolated ($J=0$) $\phi^4$-model\cite{AltlandSimons200606}. 

%Landau theory
At first, we analyse Eq. \eqref{GLW} within the Landau theory by ignoring the spatial dependence, i.e., $\phi_{\sigma}(\mib{x}) = \bar{\phi}_{\sigma}$.
Minimizing the pseudo free energy, we obtain following simultaneous equations:
\begin{eqnarray}
\label{simul}
\begin{cases}
m\bar{\phi}_{\uparrow} + J \bar{\phi}_{\downarrow} + \frac{g}{6} \bar{\phi}^3_{\uparrow} = 0 &\\
m\bar{\phi}_{\downarrow} + J \bar{\phi}_{\uparrow} + \frac{g}{6} \bar{\phi}^3_{\downarrow} = 0 &\\
\end{cases}
.
\end{eqnarray}

\begin{table}
\caption{The solutions of Eq.\eqref{simul}. We classify the solutions into 3
types according to the relation of sign between $\bar{\phi}_{\uparrow}$
and $\bar{\phi}_{\downarrow}$. (See main text for specific definitions)}
\label{solutions}
\begin{tabular}{ccc} 
\hline
Solution & Phase &Existence condition \\ 
\hline 
$\bar{\phi}_{\uparrow}=\bar{\phi}_{\downarrow}=0$ & Trivial & Always exists \\ 
$\bar{\phi}_{\uparrow}=\bar{\phi}_{\downarrow} = \sqrt[]{\frac{-6(m+J)}{g}}$ & Loop current & $J < -m $ \\ 
$\bar{\phi}_{\uparrow}=\bar{\phi}_{\downarrow} =-\sqrt[]{\frac{-6(m+J)}{g}}$ & Loop current & $J < -m $ \\ 
$\bar{\phi}_{\uparrow}=-\bar{\phi}_{\downarrow} = \sqrt[]{\frac{-6(m-J)}{g}}$ & Loop spin current & $J > m $ \\ 
$\bar{\phi}_{\uparrow}=-\bar{\phi}_{\downarrow} =-\sqrt[]{\frac{-6(m-J)}{g}}$ & Loop spin current & $J > m $ \\ 
\hline
\end{tabular}
\end{table}

Equation \eqref{simul} has five physical solutions and we classify them into three
phases: trivial, loop current, and loop-spin current as
shown in Table \ref{solutions}.
The trivial phase is characterized by the absence of the order parameters,
i.e., $\bar{\phi}_{\uparrow} = \bar{\phi}_{\downarrow} = 0$. Of course, this solution always exists. 

The loop current phase is characterized by the finite total orbital
magnetization, i.e., $\bar{\phi}_{\uparrow}+\bar{\phi}_{\downarrow} \neq 0$. The existence of the finite orbital
magnetization also means the finite loop current. Then, we call this 
phase the loop current phase. As shown in Table \ref{solutions}, the loop-current solutions exist when the condition $J < -m$ is satisfied. This phase does not possess the time-reversal symmetry, since the finite loop current breaks this symmetry. 

In contrast, the loop-spin-current phase is symmetric under the time-reversal operation, which means the absence of the total orbital magnetization,
$\bar{\phi}_{\uparrow} + \bar{\phi}_{\downarrow} = 0$. 
This condition can be achieved even in the system with the finite orbital magnetization
if the orbital magnetization from up-spin electron and that from down-spin electron have opposite directions,
$\bar{\phi}_{\uparrow} = -\bar{\phi}_{\downarrow} \neq 0$.
In this phase, the up-spin electrons and the down-spin electrons make opposite loop current, thus the electron current cancels out,
and only the spin current remains. Thus, we call this phase the loop-spin-current phase. 
As shown in Table \ref{solutions}, the loop-spin-current solutions exist if $J>m$.

%Renormalization Group
Next, the effects of the spatial fluctuations are taken into considerations via the momentum-shell RG analysis. 
For this, we perform the Fourier transformation of Eq. \eqref{GLW}, and define the scale
 transformation of the wavenumber $\mib{k}$: $\mib{k} \to \frac{\mib{k}}{b}$.
 Here, $b$ is a scale parameter. In this analysis, the momentum-shell integration are approximated as
 \begin{eqnarray}
 \int^{\Lambda}_{\frac{\Lambda}{b}} \mathrm{d}^d \mib{k} f(k) \approx S_d
 f(\Lambda) {\Lambda}^d \ln b.
 \label{shell}
 \end{eqnarray}
Here, $\Lambda$ is the cutoff momentum, $S_d$ is the surface area of a
$d$-dimensional unit sphere. 
For convenience, we rescale and nondimensionalize the parameters in Eq. \eqref{GLW} as $m/\Lambda^2 \to m$, $J/\Lambda^2 \to J $, and $\frac{S_d}{(2\pi)^d}g/\Lambda^{4-d} \to g$.
Since our one-loop level analysis is justified in the vicinity of the origin in the parameter space, these parameters should be regarded as sufficiently small.

Within one-loop perturbation, we obtain the following RG equations for dimensionless parameters up to quadratic order of small parameters:
\begin{align}
\label{rgeqst}
\frac{\mathrm{d}m}{\mathrm{d} \ln b} &= 2m + \frac{g}{2}(1-m) ,\\
\frac{\mathrm{d}J}{\mathrm{d} \ln b} &= 2J ,\\
\label{rgeqfin}
\frac{\mathrm{d}g}{\mathrm{d} \ln b} &= (4-d)g - \frac{3g^2}{2}.
\end{align}
Equations (\ref{rgeqst})-(\ref{rgeqfin}) inherit fixed points from the $\phi^4$-model since Eq. (\ref{GLW}) is identical to this model if $J=0$.
In short, the trivial Gauss fixed point, $(m, J, g) = (0, 0, 0)$, and the nontrivial Wilson-Fisher point, $(m, J, g) = (\frac{d-4}{d+2}, 0, \frac{2(4-d)}{3})$, exist. 
In this analysis, we concentrate our attention on the flow around the trivial fixed point rather than that around the nontrivial one because $m$ is assumed to be positive.
In other words, our interest exists in the fate of the flow starting from the vicinity of the origin in the parameter space.

\begin{figure}
\begin{center}
\includegraphics[scale=0.25]{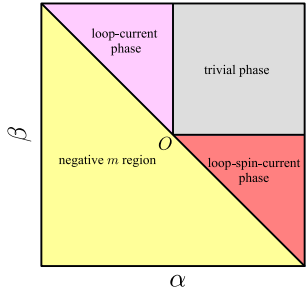}
\caption{(Color online) The schematic phase diagram for the renormalized parameters $\alpha$ and $\beta$ derived from the Landau theory. 
Since stabilities of the phases are determined only by the signs of the parameters $\alpha$ and $\beta$, 
this schematic diagram is independent of the parameter $g$. }
\label{fig:phase}
\end{center}
\end{figure}

The existence of the ordered phase is determined whether renormalized parameters satisfy the existence condition shown in Table \ref{solutions}.
In order to simplify the relation between the existence condition and the renormalized parameters, 
we introduce following valuables:
\begin{align}
\alpha = \frac{m+J}{2},  \\
\beta = \frac{m-J}{2}
.
\end{align}
With these valuables, the negative renormalized $\alpha (\beta)$ indicates the existence of the loop-current (loop-spin-current) phase 
as schematically shown in Fig \ref{fig:phase}.
Performing this transformation, the RG equations for $\alpha$ and $\beta$ are given by
\begin{eqnarray}
\label{eq:trrgeqst}
\frac{\mathrm{d} \alpha}{\mathrm{d} \ln b} = 2\alpha + \frac{g}{2}(1- \alpha - \beta), \\
\label{eq:trrgeqfin}
\frac{\mathrm{d} \beta}{\mathrm{d} \ln b} = 2\beta + \frac{g}{2}(1- \alpha - \beta).
\end{eqnarray}
The RG equation for $g$ is invariant under this transformation.
In this representation, the condition $m > 0$ is identical to $\alpha + \beta > 0$.
In order to compare our phenomenological theory and the previous studies, we set $d = 2$ hereafter.

\begin{figure}
\begin{center}
\includegraphics[scale=0.15]{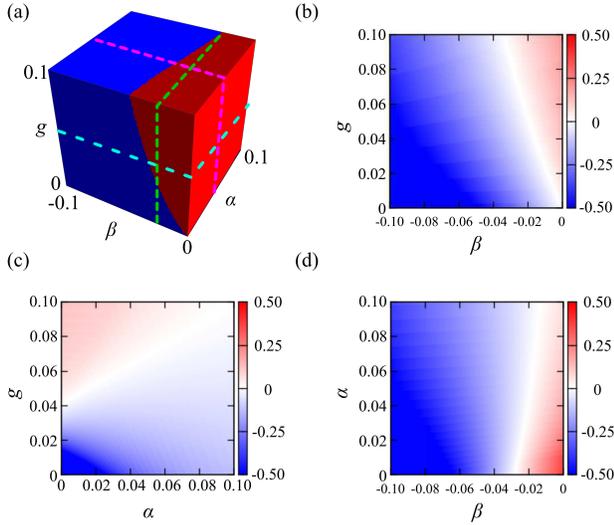}
\caption{(Color online) (a) A panoramic phase diagram for loop-spin current and trivial phases given by the RG equations. The blue (red) region corresponds to the loop-spin-current (trivial) phase. A phase is determined by the sign of the parameter $\beta$ after parameters renormalized sufficiently (See main text for the detailed definition). The dashed purple, green, and cyan lines represent the planes corresponding to a phase diagram (b), (c) and (d), respectively.  (b) A phase diagram in the $\beta$-$g$ plane with $\alpha = 0.05$. Color indicates the value of the parameter $\beta$ after the renormalization. (c) A phase diagram in the $\alpha$-$g$ plane with $\beta=-0.025$. (d) A phase diagram in the $\alpha$-$\beta$ plane with $g=0.05$.}
\label{fig:Cubic}
\end{center}
\end{figure}

Figures \ref{fig:Cubic}(a)-\ref{fig:Cubic}(d) represent phase diagrams for negative $\beta$ region given by Eqs. (\ref{rgeqfin})-(\ref{eq:trrgeqfin}).
Since Eqs. (\ref{rgeqfin})-(\ref{eq:trrgeqfin}) are valid for small parameters, 
we can discuss renormalization flows only in the vicinity of the Gauss fixed point. 
Consequently, in Figs. \ref{fig:Cubic}(a)-\ref{fig:Cubic}(d), a phase is determined by the renormalized parameters when they can be considered as small but sufficiently renormalized. 
Specifically, we consider parameters sufficiently renormalized when the absolute value of any renormalized parameter reaches the upper limit $0.5$.
When we change the upper limit value slightly, a phase diagram varies quantitatively but keeps its qualitative characters.
Thus, we should discuss only qualitative aspects of these phase diagrams.

As seen in Figs. \ref{fig:Cubic}(a) and \ref{fig:Cubic}(b), the loop-spin-current phase is stable for any negative $\beta$ and sufficiently small $g$.
This fact means that  the loop-spin-current phase is stable against small amplitude spatial fluctuations from $g$-term.
However, those figures also indicate that this phase become unstable for large $g$. 
In other words, the loop-spin-current phase is destabilized by too strong spatial fluctuations from $g$-term.
From the behavior of the stability against spatial fluctuations, we see that there exists the critical strength of fluctuations for loop-spin-current phase.

Figures \ref{fig:Cubic}(c) and \ref{fig:Cubic}(d) indicate that large $\alpha$ makes the phase more stable against fluctuations from $g$-term 
while increase the renormalized value of $\beta$. 
In short, increasing $\alpha$, which is equivalent to increasing $m$ and $J$, 
results in both positive and negative effects on the stability of the loop-spin-current phase.
In contrast, as shown in Figs. \ref{fig:Cubic}(b) and \ref{fig:Cubic}(d), decreasing $\beta$, which means decreasing $m$ and increasing $J$, gives only positive effects on the stability.
Thus, small $m$ and large $J$ make the loop-spin-current phase more stable.
Since Eqs. (\ref{rgeqfin})-(\ref{eq:trrgeqfin}) are invariant under the exchange of the parameters $\alpha$ and $\beta$,
 the above results are also obtained in the negative $\alpha$ region, which corresponds to the loop-current phase. 

%discussion

We consider that our analysis is not very reliable quantitatively but succeeds in describing the physics of the loop-current and loop-spin-current phases qualitatively. 
As is well known, the perturbative momentum-shell RG approach for the $\phi^4$-model gives quantitatively reliable results 
when the spatial dimension $d$ is close to 4\cite{AltlandSimons200606}.
For our model represented in Eq. (\ref{GLW}), the $g$-term become relevant in $d < 4$ like the $\phi^4$-model.
Thus, the same statement can be applied to our model, and the results derived from this analysis are expected to inaccurate quantitatively since we set $d=2$.
In qualitative aspects, however, the results we obtain share common features with previous studies. 
According to the numerical study for the loop-current phase on a half-filled square lattice \cite{PhysRevB.85.125117}, 
this phase has been metastable for small Hubbard interaction $U$ and become unstable for sufficiently large $U$.
Furthermore, the theoretical study based on a strong-coupling approach \cite{PhysRevB.64.220509}
has indicate the absence of the loop-current phase in the Hubbard model on cuprates.
Since the loop-current phase is unstable against too strong fluctuations of the order parameter,  it can be thought that strong quantum fluctuations destabilize the loop-current phase in these studies based on the Hubbard model.
Although the fluctuations in our analysis are not the quantum one, our analysis and those quantum theories can be thought as describing similar physics in a perspective of the fluctuation effects on the loop-current phase.

As described above, our simple theory is intrinsically classical and cannot describe the quantum fluctuations such as the Nambu-Goldstone mode.
This limitation reflects the artificial symmetry for the permutation of the parameters $\alpha$ and $\beta$ on Eqs. (\ref{eq:trrgeqst}) and (\ref{eq:trrgeqfin}).
As discussed in Ref. \citen{PhysRevLett.100.156401}, the loop-current and loop-spin-current phases have different energy because of 
the difference of the symmetries those phases break. In short, the loop-current phase breaks the discrete time-reversal symmetry, while the loop-spin-current phase breaks the continuous rotational symmetry in spin space. 
Consequently, the Nambu-Goldstone mode exists only in the loop-spin-current phase, and its zero point vibration lowers the energy of the system. 
Therefore, the loop-current phase and the loop-spin-current phase are not symmetric when the effects of the quantum fluctuations are included.

For satisfying the condition listed in Table \ref{solutions} and stabilizing loop-current phases, it is desirable that the energy cost $m$ should be small.
If the energy cost for generating the magnetization is small, itinerant electrons respond strongly to the external magnetic field. Thus, the small $m$ corresponds to the large orbital susceptibility.
Such large orbital susceptibility can be seen in the system with linear dispersion\cite{JPSJ.76.043711}.
If the Fermi level lies at the Dirac point in two dimensions, the orbital susceptibility diverges diamagnetically in the clean limit and at zero-temperature.
With such large orbital susceptibility, the loop-current or loop-spin-current phase emerges for arbitrary small $|J|$.
Indeed, the numerical calculation on a half-filled honeycomb lattice\cite{ourpapar}, where the Fermi level lies at the Dirac point, 
has shown the existence of the loop-spin-current phase within the VCA.
Therefore, the loop-current or loop-spin-current phase may be seen universally in the system with the linear dispersion as long as the Fermi level is sufficiently close to the Dirac point.

On the other hand, relationship between the macroscopic interaction parameter $J$ and microscopic interactions is not clear.
In order to see this relationship, we have to obtain the phenomenological Hamiltonian from a specific microscopic one as done in the Ginzburg-Landau theory of superconductors. 
However, a microscopic description for the orbital magnetization in periodic non-interacting systems is obtained only recently\cite{PhysRevLett.95.137205, PhysRevB.74.024408, PhysRevLett.99.197202}, and a description for periodic interacting systems is proposed only within the specific theoretical frameworks\cite{PhysRevLett.99.197202, nourafkan2014orbital}. 
Consequently, it is quite difficult to relate the parameter $J$ and microscopic interactions in periodic systems at present. 
Since the orbital magnetization in microscopic systems is not understood sufficiently, unveiling the relation between macroscopic and microscopic parameters are left for future work.

%conclusion
In conclusion, we have constructed a simple phenomenological theory for describing the loop-current and loop-spin-curent phases, and found these phases are stabilized if the interaction between the orbital magnetizations is sufficiently strong. 
Analyzing the phenomenological action via the momentum-shell RG theory, it has been shown that the phases are stable against the sufficiently small spatial fluctuations. 
These loop-current and loop-spin-current phases are expected to be seen in systems with linear dispersion, where the orbital susceptibility can diverge diamagnetically. 

\section*{Acknowledgement}
This work was supported in part by Grants for Excellent Graduate School, MEXT, Japan.

\bibliographystyle{jpsj}
\bibliography{loopsubmit.bib}

\end{document}